\documentclass[copyright]{eptcs}
\providecommand{\event}{PDMC 2009} 
\providecommand{\volume}{??}
\providecommand{\anno}{2009}
\providecommand{\firstpage}{1}
\providecommand{\eid}{??}
\usepackage{graphicx}

\title{DiVinE-CUDA -- A Tool for GPU~Accelerated~LTL~Model~Checking%
\thanks{This work has been partially supported by Czech Science Foundation grants
  No. 201/09/P497, 102/09/H042 and by Academy of Sciences of Czech Republic grant No. 1ET408050503.}}
\author{Ji\v{r}\'{\i} Barnat
\institute{Faculty of Informatics\\ Masaryk University\\ Brno, Czech Republic}
\email{barnat@fi.muni.cz}
\and
Lubo\v{s} Brim 
\institute{Faculty of Informatics\\ Masaryk University\\ Brno, Czech Republic}
\email{brim@fi.muni.cz}
\and
Milan \v{C}e\v{s}ka
\institute{Faculty of Informatics\\ Masaryk University\\ Brno, Czech Republic}
\email{xceska@fi.muni.cz}
}
\def\titlerunning{DiVinE-CUDA -- A Tool for CUDA~Accelerated~LTL~Model~Checking}
\def\authorrunning{J. Barnat \& L. Brim \& M. \v{C}e\v{s}ka}
\begin{document}
\maketitle

\begin{abstract}
  In this paper we present a tool that performs CUDA accelerated LTL Model
  Checking. The tool exploits parallel algorithm MAP adjusted to the NVIDIA CUDA
  architecture in order to efficiently detect the presence of accepting cycles
  in a directed graph. Accepting cycle detection is the core algorithmic
  procedure in automata-based LTL Model Checking. We demonstrate that the tool
  outperforms non-accelerated version of the algorithm and we discuss where 
  the limits of the tool are and what we intend to do in the future to avoid them.
\end{abstract}

\section{Introduction}
Verification and validation became an important part of the design process.
Unfortunately, the gap between the complexity of systems the current formal
verification tools can handle and the complexity of systems built in practice is
still quite wide. Therefore, any technique that accelerates the verification
process is highly desirable. A possible way to reduce the delay due to the
formal verification process is to accelerate the computation of verification
tools using contemporary parallel hardware. Hardware platforms such as
multi-core multi-cpu systems or many-core hardware accelerators, e.g. GPGPUs,
have recently received a lot of attention in this aspect.

CUDA (Compute Unified Device Architecture) is a parallel computing architecture
developed by NVIDIA~\cite{CUDA}. Recently, it has been successfully used to
accelerate formal verification process for selected settings. In~\cite{BES09}
authors demonstrated significant speedup in the verification of probabilistic
systems, while in~\cite{ES08,SOCS09} CUDA has been used to accelerate disk-based
model checking and state space generation. Let alone the CUDA technology, other
many-core hardware acceleration platforms have been tried. For example, an
implementation of FPGA accelerated Mur$\varphi$\cite{stern97parallelizing}
verification tool has been reported in~\cite{FLL08}.

In this paper we introduce a new CUDA accelerated verification tool for model
checking formulas of Linear Temporal Logic (LTL). The problem of LTL model
checking is well established problem in the formal verification
community. Computationally the problem reduces to the problem of detection of an
accepting cycle in a directed graph~\cite{vardi86anautomata}. The new tool
builds upon the DiVinE~\cite{BBC+06} framework, hence the name of the tool is
\emph{DiVinE CUDA}.

\section{DiVinE CUDA Algorithmics}
DiVinE-CUDA employs algorithm MAP~\cite{BCMS04} for accepting cycle
detection. The algorithm is, however, formulated as a repeated matrix-vector
product procedure~\cite{BBC09} in order to efficiently utilize CUDA
architecture. The idea of the MAP algorithm is as follows. Given a directed
graph with accepting vertices, the algorithm impose ordering on accepting
vertices and repeatedly computes the maximal (w.r.t. the ordering) accepting
predecessor $map(u)$ for every accepting vertex $u$ in the graph. If the
algorithm detects an accepting vertex that is its own maximal accepting
predecessor, then the vertex lies on an accepting cycle and the algorithm
terminates. In the other case, all accepting vertices that were maximal
accepting predecessors for some other vertices are marked as non-accepting
(because they do not lie on an accepting cycle) and the procedure is restarted
(goes to the next iteration). The algorithm terminates either if accepting cycle
is found or there are no more accepting vertices in the graph. From technical
reasons we employ MAP algorithm on a transposed state space graph, note that
graph transposition preserves the presence of accepting cycles.

The main computation demanding step of the algorithm is the computation of the
maximal accepting predecessor for every accepting vertex. This is done by means
of value propagation of accepting vertices along edges in the graph. If multiple
values are propagated into a single vertex, the maximum among all the incoming
values and the value of the vertex is computed and used for further
propagation. Every vertex keeps the maximum value that has been propagated
through the vertex. Once a fix-point is reached (no value can be improved),
values of maximal accepting successors are computed.

In DiVinE CUDA tool it is the maximal accepting successor computation that is
accelerated with CUDA device. In particular, relevant parts of the graph to be
analyzed are represented in an adjacency matrix. Having the matrix, the value
propagation can be realized as matrix-vector product~\cite{BBC09} for
computation of which the CUDA architecture is known to be extremely
efficient~\cite{garland}.

When initiated the DiVinE CUDA tool proceeds as follows. It starts a thread that
computes the adjacency matrix needed for CUDA processing. We use CSR (compressed
sparse row) format to store the matrix.  Note that we do not list all reachable
states in the matrix, but only those that are in components containing some
accepting vertices~\cite{lafuente02}. This feature significantly reduces the
size of the matrix to be handled. (The size reduction is up to 20-30\% of the
full size in most cases). At the same time the tool runs a second thread that
repeatedly performs CUDA accelerated accepting cycle detection on the part of
the matrix that has been computed so far. If an accepting cycle is present in
that part of the graph it is discovered before the full state space is
generated. Therefore, DiVinE CUDA works on-the-fly.

\section{Using the Tool}
DiVinE-CUDA is a tool that stems from parallel and distributed LTL Model Checker
DiVinE~\cite{BBC+06,BBR08}. As such, DiVinE-CUDA tool uses DiVinE native
modeling language DVE~\cite{BBC+06}. In DVE modeling language the system to be
verified is given as an asynchronous network of communicating finite
automata. Transitions of every automaton in the network can be augmented with
guards, buffered and unbuffered channel communication primitives, and variable
updates.

\begin{figure}[t]
  \centering
  \includegraphics[width=\linewidth]{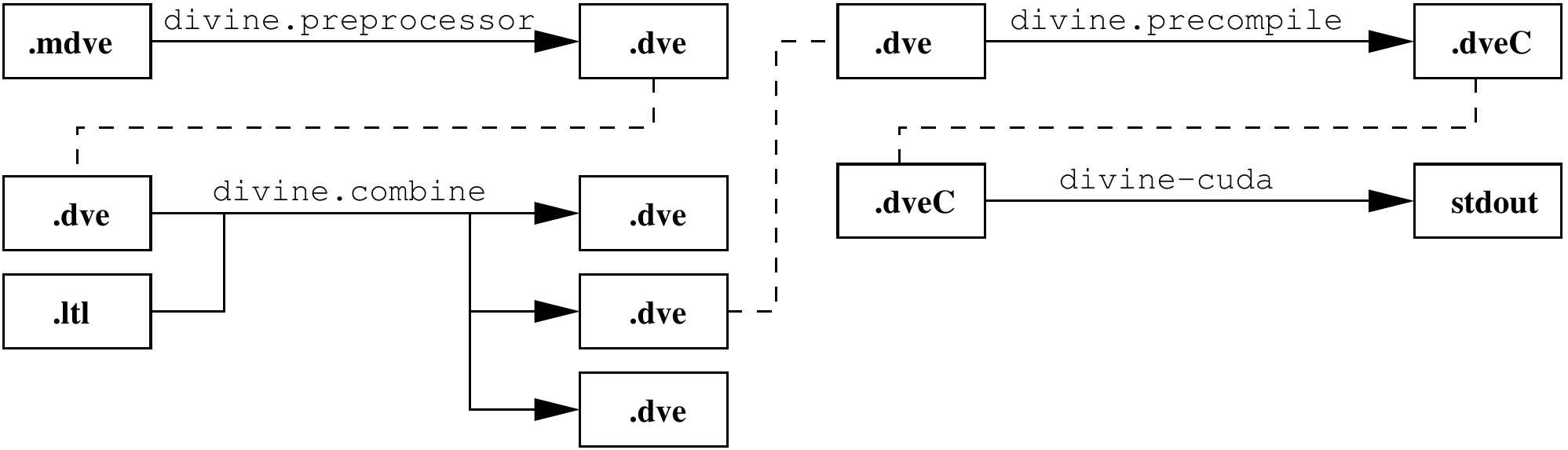}
  \caption{DiVinE CUDA work-flow.}
  \label{fig:workflow}
\end{figure}

The scheme of how the DiVinE CUDA tool should be used is given in
Figure~\ref{fig:workflow}. Having prepared the model either directly as a
\emph{.dve} file or from a \emph{.mdve} template using
\texttt{divine.preprocessor} the user has to specify the property to be
verified. The property can be given either directly as a property automaton
(also known as never claim automaton) in the model file, or as (a set of) LTL
formula(s) in a separate file, in which case the files have to be further
processed by \texttt{divine.combine} tool to get a model file with the property
automaton.

The next step in the verification process is to
produce precompiled version of the model using \texttt{divine.precompile}
tool. Precompiled version of the model (file with extension \emph{.dveC}) is
actually a dynamically linked library containing functions to generate states of
the model with specification. Finally, the precompiled representation of the
model is used as an input for the \texttt{divine-cuda} tool itself.

During the computation the tool reports periodically the numbers of generated
states and transitions, numbers of MAP iterations and CUDA device calls made so
far to the standard output. At the end the tool outputs whether an accepting
cycle has been found, in which case the given model does not satisfy the
specification, or whether no accepting cycle has been discovered, i.e. the
specification is satisfied.

\section{Experiments}
To briefly evaluate our tool we compared our implementation of CUDA accelerated
MAP algorithm with the existing algorithms implemented in the DiVinE-Cluster
version 0.8.2 model checker. For the comparison we used selected DiVinE native
models including leader election protocol, elevator cabin system, Peterson's and
Anderson's solutions to mutual exclusion problem and dining philosophers. We
tested both the models with specification error (with an accepting cycle) as
well as models without a specification error. All the experiments were run on a
Linux workstation equipped with two AMD Phenom(tm) II X4 940 Processors @ 3MHz,
8 GB DDR2 @ 1066~MHz RAM and NVIDIA GeForce GTX 280 GPU with 1GB of GPU~memory.
 
Table~\ref{fig:time} provides details on run-times of individual algorithm
parts. As for the CUDA MAP algorithm, the total run-time includes the
initialization time (not reported in the table), CSR construction time
(\emph{CSR time}), and time spent on CUDA computation (\emph{CUDA time}).  Note
that the first iteration of CPU MAP is actually slower than construction of the
CSR representation. This is because the first iteration of the CPU MAP not only
generates the state space, but also computes first stable values of \emph{map}.
Just for curiosity we also compare the performance of the new tool with DiVinE
Cluster tool running OWCTY Algorithm~\cite{Cerna03}. Algorithms MAP and OWCTY
were running on a single core.

Table~\ref{fig:comparison} gives a comparison of overall run-times for both
valid and invalid model checking instances. Though, the overall speedup is not
that significant, it is still impressive. We can also see that the burden of
data preparation is huge compared to the CUDA processing itself. 

\begin{table*}[!p]
\begin{center}
\setlength{\tabcolsep}{4pt}
\renewcommand{\arraystretch}{1.1}
\begin{tabular}{|c||c||c|c|c||c|c|c|c||c|c|} \hline
        & & \multicolumn{3}{|c||}{CUDA MAP} & \multicolumn{4}{|c||}{CPU MAP}  & \multicolumn{2}{|c|}{CPU OWCTY}\\ \cline{3-11}
        & \emph{accepting} & \emph{CSR}     & \emph{CUDA} & \emph{total} & \emph{1st iter.} & \emph{other iter.} & \emph{total} & & \emph{reachability} & \emph{total} \\[-1.5ex]
        Model   & \emph{cycle}   & \emph{time}     & \emph{time}   & \emph{time} & \emph{time} & \emph{time}  & \emph{time} & \# \emph{iter.} &\emph{time}   & \emph{time} \\ \hline 
        elevator 1    & N  &  26   & 7   & 34   & 44   &56    & 100   & 16   & 24     & 41    \\ \hline
        leader        & N  &  87   & 1   & 90   & 97   &600   & 697   & 17   & 90     & 297   \\ \hline
        peterson 1    & N  &  105  & 6   & 113  & 175  &270   & 445   & 16   & 110    & 188   \\ \hline
        anderson      & N  &  31   & 7   & 39   & 64   &51  & 115   & 5    & 33     & 113   \\ \hline \hline
        elevator 2    & Y  &  33   & 1   & 35   & 50   & --    & 50    & 1    & 41     & 177   \\ \hline
        phils         & Y  &  45   & 1   & 47   & 295  & 102   & 397   & 5    & 180    & 576   \\ \hline
        peterson 2    & Y  &  25   & 5   & 31   & 173  & --    & 173   & 1    & 114    & 404   \\ \hline
        bakery        & Y  &  24   & 1   & 26   & 240  & --    & 240   & 1    & 219    & 907   \\ \hline
\end{tabular}
\caption{Comparison of run-times (in seconds) for CUDA accelerated MAP
  algorithm, non-accelerated MAP algorithm and OWCTY algorithm.}
\label{fig:time}
\renewcommand{\arraystretch}{1}
\end{center}
\end{table*}

\begin{table*}[!p]
\begin{center}
\setlength{\tabcolsep}{4pt}
\renewcommand{\arraystretch}{1.1}
\begin{tabular}[b]{|c||c||c|c||c|c|} \hline
                      &  CUDA MAP          & \multicolumn{2}{|c||}{CPU MAP}                    & \multicolumn{2}{|c|}{CPU OWCTY}\\ \cline{2-6}
        Models        & \emph{total time}  & \emph{total time} & \emph{CUDA MAP speedup}
        & \emph{total time} & \emph{CUDA MAP speedup} \\ \hline 
        non-accepting &  276               & 1357              & 4.92                        & 639               &  2.32                 \\ \hline
        accepting     &  139               & 860               & 6.19                        & 2064              &  14.87                 \\ \hline \hline
        both          &  415               & 2173              & 5.24                        & 2730              &  6.51                 \\ \hline
       \end{tabular}
\caption{The overall run-times in seconds, and speedup of the whole model checking procedure.}
\label{fig:comparison}
\renewcommand{\arraystretch}{1}
\end{center}
\end{table*}

\begin{table}[!p]
\begin{center}
\setlength{\tabcolsep}{4pt}
\renewcommand{\arraystretch}{1.1}
\begin{tabular}{|cc|c|c|c||c|c|c||c|c|c|} 
  \hline
        & & \multicolumn{3}{|c||}{CUDA MAP} & \multicolumn{3}{|c||}{}  & \multicolumn{3}{|c|}{}\\ \cline{3-5}
        & & \emph{CSR}     & \emph{CUDA} & \emph{total} &
        \multicolumn{3}{|c||}{CPU MAP} & \multicolumn{3}{|c|}{CPU OWCTY}  \\[-1.5ex]
          &  & \emph{time}     & \emph{time}   & \emph{time}
        &\multicolumn{3}{|c||}{} & \multicolumn{3}{|c|}{} \\ \hline 

        \hline

       \multicolumn{2}{|l}{\textbf{1 core}:} &
       \multicolumn{2}{c}{ 386 + 29 = } & 415 &
       \multicolumn{2}{c}{Total time:} & \multicolumn{1}{c||}{2 173} &
       \multicolumn{2}{c}{Total time:} & 2 730

       \\

       \multicolumn{2}{|c}{} & 
       \multicolumn{3}{c||}{} &
       \multicolumn{2}{c}{Speedup:} & \multicolumn{1}{c||}{\textbf{5.24}} &
       \multicolumn{2}{c}{Speedup:} & \textbf{6.51} \\

       \hline

       \multicolumn{2}{|c}{\textbf{2 cores}:} &
       \multicolumn{2}{c}{ 193 + 29 = } & 222 &
       \multicolumn{2}{c}{Total time:} & \multicolumn{1}{c||}{1087} &
       \multicolumn{2}{c}{Total time:} & 1365

       \\

       \multicolumn{2}{|c}{} & 
       \multicolumn{3}{c||}{} &
       \multicolumn{2}{c}{Speedup:} & \multicolumn{1}{c||}{\textbf{4.87}} &
       \multicolumn{2}{c}{Speedup:} & \textbf{6.15} \\

       \hline

       \multicolumn{2}{|c}{\textbf{4 cores}:} &
       \multicolumn{2}{c}{ 97 + 29 = } & 126 &
       \multicolumn{2}{c}{Total time:} & \multicolumn{1}{c||}{544} &
       \multicolumn{2}{c}{Total time:} & 683

       \\

       \multicolumn{2}{|c}{} & 
       \multicolumn{3}{c||}{} &
       \multicolumn{2}{c}{Speedup:} & \multicolumn{1}{c||}{\textbf{4.32}} &
       \multicolumn{2}{c}{Speedup:} & \textbf{5.42} \\

       \hline

       \multicolumn{2}{|c}{\textbf{8 cores}:} &
       \multicolumn{2}{c}{ 49 + 29 = } & 78 &
       \multicolumn{2}{c}{Total time:} & \multicolumn{1}{c||}{272} &
       \multicolumn{2}{c}{Total time:} & 342

       \\

       \multicolumn{2}{|c}{} & 
       \multicolumn{3}{c||}{} &
       \multicolumn{2}{c}{Speedup:} & \multicolumn{1}{c||}{\textbf{3.48}} &
       \multicolumn{2}{c}{Speedup:} & \textbf{4.38}      
       \\\hline
\end{tabular}
\caption{A hypothetical speedup of DiVinE CUDA w.r.t. multicore parallel
  algorithms. We suppose optimal (linear) speed-up for both parallel algorithms
  MAP and OWCTY and for the CSR construction phase of the CUDA MAP algorithm.}
\label{fig:multicore}
\renewcommand{\arraystretch}{1}
\end{center}
\end{table}

\section{Availability and Future Work}
At the moment the tool cannot handle models for which the corresponding
\emph{reduced} matrix of the graph does not fit the memory of a single CUDA
device, it lacks the ability of counterexample generation, and cannot employ
multiple threads to compute the CSR representation in parallel. We intend to
address all these issues in the next version of the tool. As for the run-times,
we expect significant improvement due to parallel preparation of the CSR graph
representation. See Table~\ref{fig:multicore}. As for the limit on the size of
the verification problem, we plan to introduce sort of clever swapping mechanism
of the matrix stored in the GPU memory and to extend the memory available by
employ multiple CUDA devices.


DiVinE CUDA tool is freely available from DiVinE web pages
\footnote{\url{http://divine.fi.muni.cz/page.php?page=divine-cuda}} where we
provide both download and install instructions as well as simple tutorial on
using the tool.

\bibliographystyle{eptcs} 
\bibliography{pdmc2009}
\end{document}